# FACTORS THAT DETERMINE CONTINUOUS INTENTION TO USE MOBILE PAYMENTS IN MALAWI

Jones Ntaukira, Malawi University of Science and Technology, jones.ntaukira@zuwaenergymw.com

Priscilla Maliwichi, Malawi University of Science and Technology, pmaliwichi@must.ac.mw

James Kamwachale Khomba, Malawi University of Science and Technology, jkhomba@must.ac.mw

**Abstract:** The proliferation of mobile phones has made mobile payments to be widely used in developing economies. However, mobile payment usage in Malawi is low, and there are many limitations to encourage users to continuously use mobile payments. The purpose of this research was to examine determinants of continuous intention to use mobile payments in Malawi. A conceptual framework adapted from Technology Acceptance Model was developed. Data was collected through a survey while data analysis used Structural Equation Modelling Partial Least Squares using SmartPLS software. The findings of this study showed that society norms significantly influence continuous intention to use mobile payments ($p$=0.012). Most interestingly, prior knowledge (p=0.000) and seamlessness (p=0.000) had the strongest influence as compared to structural assurance (p=0.008). Seamlessness significantly influenced satisfaction ($p$=0.002) and society norms ($p$=0.001). Seamlessness and service quality had significantly negative effects on satisfaction. The findings of this research provide several considerations to guide the mobile payments industry in Malawi. The findings may also improve the existing mobile payments system's business models, marketing strategies, customer engagement on security issues, transparency, and interoperability of payment systems. Regulators may also find the findings of this study very insightful in advancing the mobile payments agenda in Malawi.

## 1. INTRODUCTION

Electronic commerce's emergence and success have led to the digitization of payment systems, referred to as electronic payments or e-payments (Casado-Aranda, Liébana-Cabanillas, & Sánchez-Fernández, 2018). E-payment is defined as the buying and selling of goods and services on the web (Tiwari, Buse, & Herstatt, 2006). The United States Bureau of the Census has defined e-payment as any payment completed over a computer-mediated network that involves the transfer of ownership or rights to use goods or services using free software or mobile application (as cited by Tiwari et al., 2006, p.39). However, various researchers refer to e-payment as all payments that are initiated, processed, and received electronically (Ahmed & Ali, 2017). Sometimes, it can generally be referred to as digital payments and may be linked or not be linked to a financial institution or a bank (Ahmed & Ali, 2017; Diniz, Porto de Albuquerque, & Cernev, 2011). The most common e-payments systems in the world are credit cards, debit cards, and email-based PayPal transactions (Casado-Aranda et al., 2018). However, due to the widespread adoption and usage of mobile phones worldwide, mobile commerce has emerged, giving a new dimension to e-commerce (Casado-Aranda et al., 2018).

Mobile commerce (m-commerce) uses mobile payments to pay for goods and services (Diniz et al., 2011). There are currently two kinds of mobile payment systems in use around the world: telecommunications-led mobile payment platforms and bank-led mobile payments (Madise, 2014). Comparing the two types of mobile payment platforms in Malawi, telecom-led mobile money platforms have become a key financial player (Madise, 2014). Telecom-led mobile payment is popular among the millions of people who do not use banks in Malawi, particularly in rural areas (Madise, 2014; Reserve Bank of Malawi [RBM], 2017). With commercial banks concentrating in urban and semi-urban areas, rural areas remain highly financially secluded (Madise, 2014). Mobile payments that are telecommunications-led are also referred to as mobile money, an electronic currency for transactions (Diniz et al., 2011).

Mobile money is a fast-growing industry, not only in Malawi but also in the developing world (Madise, 2014). However, despite an increase in mobile payment solutions in Malawi, data on usage suggests that very few people are using these services (Madise, 2014; RBM, 2017). In their Monthly National Payment Systems Report for July 2017, RBM reported that of the 4 million subscribers of telecom-led mobile payments, only 22.1% of registered users were active during the previous 30 days (RBM, 2017). Thus, it was due to this background that this study sought to investigate factors that determine continuous intention to use mobile payments in Malawi. Several studies have examined factors that determine continuous intention to use mobile payments (Agwu, 2017; Ahmed & Ali, 2017; Onsongo & Schot, 2017). However, little is known of the factors that determine continuous intention to use mobile payments in Malawi (Madise, 2014; Nyirenda & Chikumba, 2014; Tsilizani, 2015). Therefore, the question which guided this study was:

● What factors determine the continuous intention to use mobile payments in Malawi?

The study used a conceptual framework to understand the factors determining people's continuous intention to use mobile payments in Malawi. Based on the findings of this study, we then laid out considerations that may assist the mobile payment service industry in Malawi on how they could make mobile payments in Malawi to be used at its fullest potential.

## 2. Literature Review

This section discusses theoretical models and literature on continuous intention to use technologies, which is a post-adoption behaviour. To understand continuous intention to use technologies, the researchers suggested and employed several models and theories. Some of these were extended Technology Acceptance Model and IS Impact Model. The study also adopted constructs from other theoretical models implemented to extend Technology Acceptance Model (TAM) to understand key determinants of continuous intention to use technologies.

One major flaw of TAM model is its inability to offer actionable guidance to developers (Bradley, 2009; Park, 2009). UTAUT, as an updated TAM model, has also received various criticism, some of which is its complexity (van Raaij & Schepers, 2008) and 'chaotic' (Bagozzi, 2008), hence the need to stick to simple models with proper screening procedures. This study, therefore, used Extended TAM and incorporated concepts from IS Impact model to explain continuous intention. By incorporating constructs from IS Impact Model, the researchers aim to address a concern of understudying usage in TAM (Bradley, 2009) while also using Task Technology Fit (TTF) (a construct in IS Impact Model) to explore actionable guidance for developers. The following sections discuss the genealogy of the constructs used in this study.

### 2.1. Extended Technology Acceptance Model (TAM)

Technology Acceptance Model (TAM) was developed by Davis, Bagozzi, & Warshaw (1989). Davis, Bagozzi, & Warshaw (1989) found two main reasons people adopt or reject a technology, namely perception of the technology usefulness and perception of ease of use of the technology. TAM is one of the most widely used model when predicting information technology adoption, and according to different scholars, it consistently explains between 40% to 50 % of user acceptance and actual usage (Surendran, 2012; Venkatesh & Bala, 2008).

TAM was then extended to include a possible explanation of *subjective norm*, perceived usefulness and usage intentions (Venkatesh & Davis, 2000). Later, TAM was extended by Rose & Fogarty (2006) to include *perceived risk* as one of the construct as shown in Figure 1.

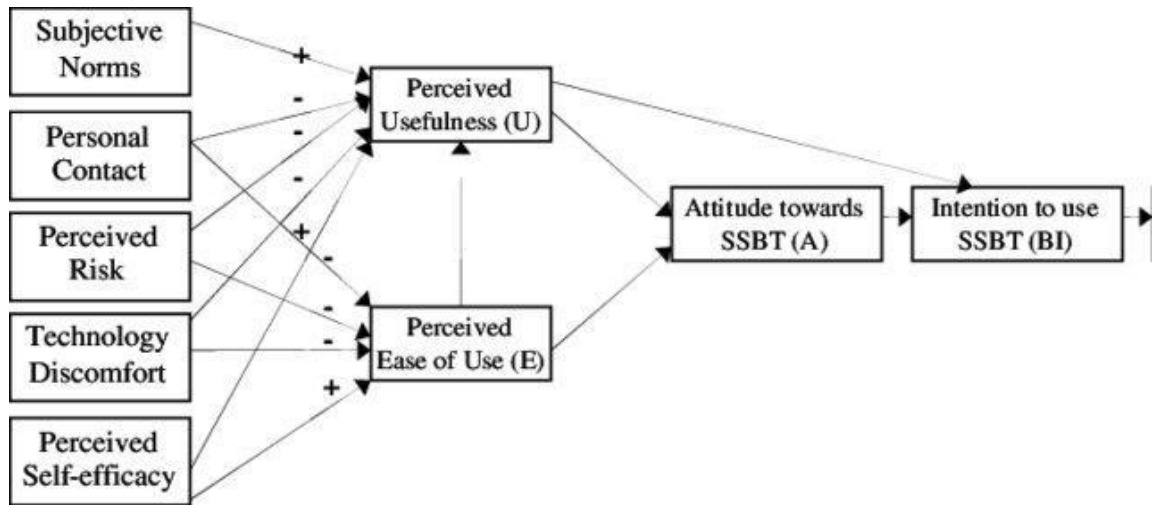

**Figure 1. Extended Technology Acceptance Model -Adapted from** Rose & Fogarty (2006)

### 2.1.1. Perceived Usefulness, Perceived Ease of Use and Attitude Towards Intention to Use Technologies

*Perceived usefulness* and *perceived ease of use* were original constructs of TAM and are key determinants of acceptance of an information system (Davis et al., 1989). *Perceived usefulness* is the degree to which consumers believe that a particular technology will facilitate the transaction process. While, *perceived ease of use* is the degree to which a consumer believes that using a particular technology will be effortless. Applying these constructs in mobile commerce, in Nigeria, it found that electronic banking provides a higher degree of convenience that enables customers to access their money at all times and places (Agwu, 2017). Perceived usefulness directly *influence attitude or intention to use*, but perceived ease of use acts indirectly through usefulness (Pavlou, 2003).

### 2.1.2. Perceived Risk, Service Quality and System Quality versus Satisfaction

Over the years, researchers have extended TAM to better understand attitudes. In 2005, Wixom and Todd developed a model using construct from TAM, theory of reasoned action (TRA) and unified theory of acceptance and use of technology (UTAUT) to understand beliefs and attitudes when using a system. In their model *satisfaction* and *system quality* were some of the constructs of their model (Wixom & Todd, 2005). Later, the model was extended to include *service quality* by Xu, Benbasat and Cenfetelli (2013).

*Satisfaction* has been identified as a major predictor of behavioral intention to use mobile payments (Ahmed & Ali, 2017; Wixom & Todd, 2005). To measure satisfaction, other factors must be used to explain the major predictor. Several prior studies used perceived risk, service quality, and system quality as antecedent factors of satisfaction (Agwu, 2017; Ahmed & Ali, 2017). Generally, consumers associate risk with loss of money; hence, security and privacy are highly significant predictors of adoption (Agwu, 2017). In contrast, Sinha (2010) found that *perceived risk* was not significant in online shopping, but at the gender level, they found that men and women behaved differently due to *perceived risk*. Therefore, *service quality* and *system quality* are antecedent factors that significantly determine the continuous intention to use mobile money transfer (Ahmed & Ali, 2017).

## 2.2. IS Impact Model

IS Impact Model was developed by Ahmed and Ali (2017) to understand the determinants of continuous intention to use mobile money payments technologies. IS Impact Model is a combination of Task Technology Fit model, Theory of Reasoned Action, Technology Acceptance Model and Information Systems Success Model (Ahmed & Ali, 2017). In this study, the researchers adopted trust, task technology fit, firm reputation, structural assurances and continuous intention constructs from the integrated model.

### 2.2.1. Satisfaction and Continuous Intention

Literature suggests that one of the key determinants of *continuous intention* to use technologies is *customer satisfaction* (Susanto, Chang, & Ha, 2016). A satisfied customer is more likely to continue using a service or product. In Information Systems (IS) field, user satisfaction with technology is also vital in influencing the adoption and continuance use of technology (Ahmed & Ali, 2017; Susanto et al., 2016). Ahmed and Ali (2017) reported a significant positive relationship between satisfaction and continuous intention to use mobile money transfer. They concluded that satisfied customers are more likely to continue using a mobile money transfer.

### 2.2.2 Trust and Continuous Intention

User's *trust* towards mobile payments impacts their continuous behaviour intention to use technologies (Pavlou, 2003; Susanto et al., 2016). Trust defined as a belief that the other party will behave in a socially responsible manner and, by so doing, will fulfil the trusting party's expectations without taking advantage of their vulnerabilities (Pavlou, 2003). Several studies have found that trust is positively related to continuance intention to use technologies (Dupas, Green, Keats, & Robinson, 2016; Susanto et al., 2016). In a study to understand the challenges of mobile banking in Kenya, it was found that, out of the 63% of people who opened an account, only 18% actively used it (Dupas et al., 2016). Survey evidence from the study suggested that people did not begin saving in their bank accounts because they did not trust the bank (Dupas et al., 2016).

### 2.2.3. Subjective Norms and Continuous Intention

*Subjective norms*, also called social factors, may negatively affect users' continuous behaviour to use mobile payments (Ahmed & Ali, 2017). Subjective norm is defined as the perceived social pressure to perform or not perform the behaviour (Ahmed & Ali, 2017). In India, the lower usage of online shopping was highly attributed to social-psychological factors influence of friends and relatives (Sinha, 2010).

Similarly, others have also found that the opinion of one's surroundings, including friends, colleagues, and family members, significantly contributed to one's adoption of mobile money transfer (Ahmed & Ali, 2017). Ahmed & Ali (2017) specifically found that subjective norm was the best predictor of continuous intention to use mobile money transfer among their study respondents (Ahmed & Ali, 2017).

### 2.2.4. Task Technology Fit versus Perceived Usefulness

*Task Technology Fit* is another antecedent factor that this research introduced to further our understanding of mobile payments' perceived usefulness, especially for corporate users and SMEs. The researchers were curious to find out if, at all, a better fit between task requirements and mobile payment system functionalities increases perceived usefulness. Others have argued that task technology fit may impact perceived usefulness (Ahmed & Ali, 2017).

### 2.2.5. Firm Reputation and Structural Assurances versus Trust

*Trust* is a defining feature of most economic and social interactions in which uncertainty is present (Pavlou, 2003). Trust has always been an important element in influencing consumer behavior (Pavlou, 2003). Several studies have found that trust is a major determinant of intention to adopt or continue using mobile banking services among different users (Ahmed & Ali, 2017; Lin & Wang, 2006). For example, a study in Iran found that customer's intention to adopt mobile banking was significantly influenced by how much trust the customers exert on the service provider (Hanafizadeh, Behboudi, Abedini Koshksaray, & Jalilvand Shirkhani Tabar, 2014). Hence, *firm reputation* and *structural assurance* are antecedents that addresses institutional based trust (Ahmed & Ali, 2017). It has been urged that trust has a significant positive impact on customer satisfaction and loyalty (Lin & Wang, 2006).

## 3.  CONCEPTUAL FRAMEWORK

The conceptual framework used in this study is a combination of extended Technology Acceptance Model and other adoption models based on TAM, and integrated model which test continuous intention to use technologies.

### 3.2.  Seamlessness

Seamlessness is a construct that the current study introduced to understand user behaviour towards mobile payments in Malawi. In this study, seamlessness is defined as the technical and operational compatibility between two different mobile payment systems regardless of whether the users are on the same network or not.

### 3.3.  Constructs on Determinants of Continuous Intention to Use Technologies

Based on the constructs discussed in the literature review section and a construct introduced in this section (seamlessness), Figure 2 summarises the hypothetical model that was developed for this study.

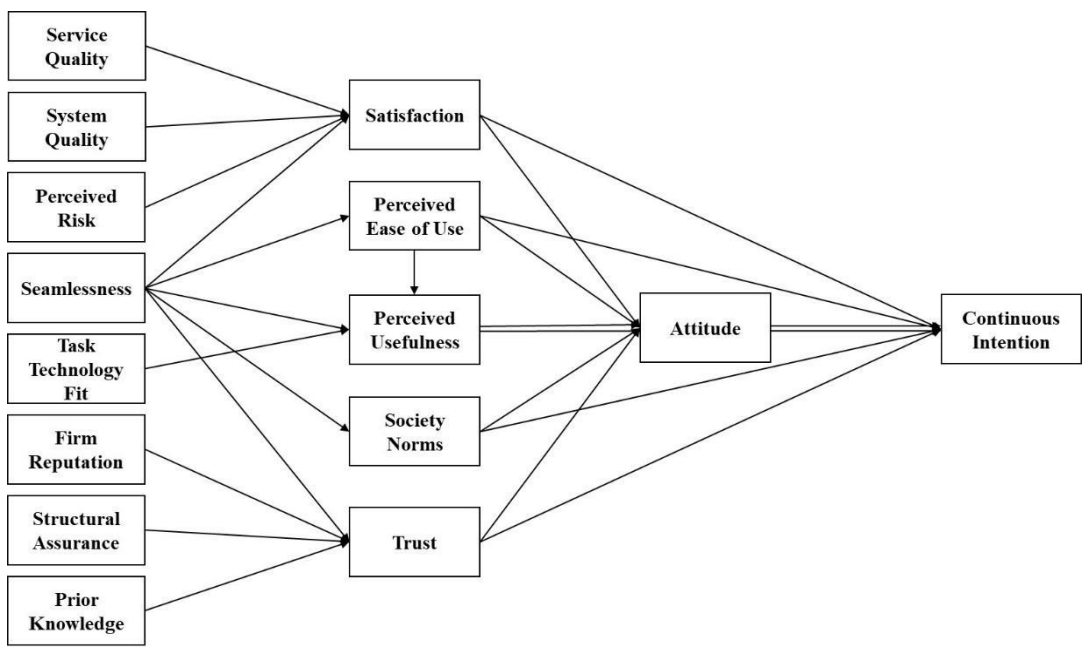

**Figure 2. Hypothesised Model for Determinants of Behaviour for Using Mobile Payments in Malawi.**

A list of hypothesises used in this study based on figure 2 are listed in Table 1.

| No | Description |
| --- | --- |

| H1a | Satisfaction with mobile payment systems positively determines the user's continuous intention behaviour to use mobile payments. |
|---|---|
| H1b | Satisfaction with mobile payments systems positively determines user's attitude on the use of mobile payments |
| H2a | User's trust towards mobile payments systems and companies positively determine continuous intention behaviour to use mobile payments technology |
| H2b | User's trust towards mobile payments positively determines their attitude on the use of mobile payments |
| H3a | Perceived usefulness of mobile payments positively determines continuous intention behaviour to use mobile payments technology |
| H3b | Perceived usefulness of mobile payments positively determines user's attitude on the use of mobile payments |
| H4a | Societal norms towards mobile payments positively determines user's continuous intention behaviour to use mobile payments |
| H4b | Societal norms towards mobile payments positively determines user's attitude on using mobile payments |
| H5a | Perceived ease of use of mobile payment technology has a positive impact on continuous intention to use mobile payments |
| H5b | Perceived ease of use of mobile payment technology has a positive impact on user's attitude on the use of that mobile payment technology |
| H5c | Perceived ease of use of a mobile payment technology has a positive impact on the perceived usefulness of mobile payments. |
| H6 | Attitude of users of mobile payments towards the use of the same positively determines their continuous intention behaviour to use mobile payments |
| H7a | Seamlessness of mobile payment technology has a positive impact on perceived usefulness of mobile payments |
| H7b | Seamlessness of mobile payment technology has a positive impact on perceived ease of use of mobile payments |
| H7c | Seamlessness of mobile payment technology has a positive impact on society norms of mobile payments |
| H7d | Seamlessness of mobile payment technology has a positive impact on user's trust in mobile payments |
| H7e | Seamlessness of mobile payment technology has a positive impact on user's satisfaction with mobile payments |
| H8a | Prior knowledge of mobile payment technology has a positive impact on society norms on mobile payments |
| H8b | Prior knowledge of mobile payment technology has a positive influence on user's trust in mobile payments |
| H9a | The perceived risk that user's associate with mobile payments has a positive influence on their satisfaction with mobile payments |
| H9b | The perceived risk that user's associate with mobile payments has a positive influence on their trust in mobile payments |
| H10 | Service quality of mobile payments providers has a positive influence on user's satisfaction with mobile payments |
| H11 | System quality of a mobile payments platform has a positive influence on user's satisfaction with mobile payments user's satisfaction |
| H12 | User's perception of a mobile payment technology fitting their tasks and roles has a positive influence on user's perceived usefulness |

| H13a | Structural assurances that service providers give to users of mobile payments have a positive influence on their society norms on the use of mobile payments |
|---|---|
| H13b | Structural assurances that service providers give to users of mobile payments have a positive influence on user's trust in the use of mobile payments |
| H14 | A mobile payments firm's reputation has a positive impact on user's trust in the use of mobile payment |

**Table 1. List Hypothesises Based on the Conceptual Framework**

# 4. METHODOLOGY

This study used quantitative research methods. Data was collected using a survey approach to understand the determinants of continuous intention to use mobile payments in Malawi by current users. The questionnaire was developed based on the conceptual framework discussed in Section 2. Data analysis used Partial Least Squares Structured Equation Modelling (PLS-SEM) using SmartPLS3 software and SPSS (version 22). Structural Equation Modelling (SEM) is a second-generation multivariate data analysis method used in behavioral sciences to test theoretically supported linear and additive causal models (Wong, 2013). On the other hand, PLS is a soft modelling approach to SEM, which makes no assumptions about data distribution (Vinzi, Chin, Henseler, & Wang, 2010).

## 4.1. Research Design

The specific quantitative strategy that was used for this study was a survey design. Survey research provides a quantitative or numeric description of trends, attitudes, or opinions of a population by studying a population sample (Creswell, 2009). It can be used in cross-sectional and longitudinal studies using questionnaires or structured interviews for data collection to generalize results from a sample to a population (Creswell, 2009). The survey design was preferred for this study for many reasons. Firstly, a small sample of mobile payments users would need to be studied in order to make conclusions about the population other than studying the whole population. Secondly, survey research is economical and does not require many resources if it is done effectively. Thirdly, there is a rapid turnaround in data collection. The survey conducted for this study was cross-sectional, whereby data was collected at one point in time rather than spread over a long period of time.

## 4.2. Setting and Population

The study was conducted in Lilongwe, covering both Lilongwe city and Lilongwe rural. Data was collected from mobile payment service users, both telecom-led and bank-led services between end January 2018 and February 2018. The population for this study was defined as subscribers who have an account with any mobile payments service providers. Inclusion criteria comprised the following: above 18 years of age, male or female, able to read and write English, must have used their mobile payment service in the last six months.

## 3.3. Sampling frame and Sample size

According to Jackson (2009), a sampling frame is the set of people who has a chance to be selected given the chosen sampling approach. Because the population size was greater than 50,000 (actual population was 4, 581,244), the formula for calculating a sample for infinite populations was used as proposed by Godden (2004). Thus, a formula was used to calculate sample size with 90% confidence level and a probability of 5% with a margin of error of 4% as given in Godden (2004). Hence, our sample was 422 respondents.

### 3.4. Data Analysis

Data was coded and entered into Microsoft Excel before exporting it into Statistical Package for Social Scientists software (SPSS) for descriptive analysis (i.e. mean, standard deviation, variance, frequency, percent, and correlation). To understand the relationships between the constructs, the constructs were first tested for validity and reliability using Exploratory Factor Analysis (Ahmed & Ali, 2017). After confirming the reliability and validity of the constructs, structural model analysis followed. Structural model analysis was done following a procedure for Partial Least Squares Structural Equation Modeling using SmartPLS 3 software, as shown in in Figure 3.

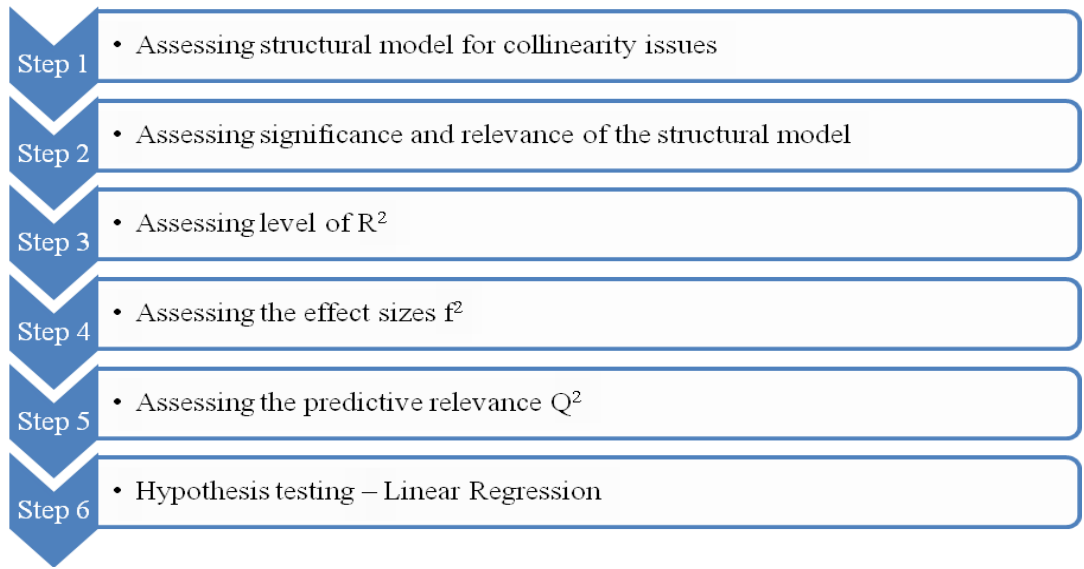

**Figure 3. Step by step Structural Model Assessment Procedure for this study - Adapted from Hair, Hult, Ringle, & Sarstedt (2014)**

## 5. RESULTS

The study distributed 422 questionnaires, but only 400 questionnaires were returned, representing a response rate of 95% of the original sample size. From the 400 returned questionnaires, seven questionnaires were not used because three respondents said they had not used mobile money before. Four respondents only partially answered the questions or some pages were missing from their questionnaires. Therefore 393 questionnaires were used for data analysis representing a final response rate of 93%.

### 5.1. Demographic Characteristics of the Respondents

This section presents the demographic characteristic of the respondents. These characteristics are based on demographic attributes, as well as the uses of mobile payment services.

#### 5.1.1. Demographic Characteristics of Respondents

Women represented 60.3% of the respondents, while 39.4% were men and 0.3% did not want to disclose their gender. By age, the age bracket of 21 to 30 had the highest respondents at 57%. More than 80% of the respondents had at least a Malawi School Certificate of Education (MSCE) as minimum education qualification. In terms of marital status, most respondents (66.7%) were single. Table 2 summarises the demographic characteristics of the respondents.

| Variable (n=393) | Category | Frequency | Percentage |
|---|---|---|---|
| Gender | Male | 155 | 39.4 |
|  | Female | 237 | 60.3 |
|  | Rather not disclose | 1 | 0.3 |
| Age | Less than 20 | 69 | 17.6 |
| | 21-30 | 224 | 57.0 |
| | 31-40 | 59 | 15.0 |
| | 41-50 | 32 | 8.1 |
| | >50 | 9 | 2.3 |
| Education | Primary school | 11 | 2.8 |
| | Secondary (MSCE) | 132 | 33.6 |
| | Certificate/Diploma | 128 | 32.6 |
| | Bachelors degree | 113 | 28.8 |
| | Postgraduate | 9 | 2.3 |
| Marital Status | Married | 119 | 30.3 |
| | Single | 262 | 66.7 |
| | Widow/Widower | 9 | 2.3 |
| | Divorced | 0 | 0.0 |
| | Separated | 3 | 0.8 |
| Head of household | Head of household | 139 | 35.4 |
| | Not Head of house | 254 | 64.6 |

**Table 2. Demographic Characteristics of Respondent**

### 5.1.2. Uses of Mobile Payment Services

The findings of this study have shown that 58.5% of the respondents used mobile payment services for sending money to others, while 24.9% used mobile payment services for buying airtime on their mobile phones. Only 9.4% of the respondents used mobile payment services for shopping in stores and shops. Table 3 summarises uses of mobile payment services in Malawi.

| Variable (n=393) | Category | Frequency | Percentage |
|---|---|---|---|
| Use of mobile payments | Buying from shops/stores | 37 | 9.4 |
| | Sending money to others | 230 | 58.5 |
| | Buying airtime | 98 | 24.9 |
| | Paying school fees | 5 | 1.3 |
| | Paying utility bills | 12 | 3.1 |
| | Buying things online | 3 | 0.8 |
| | Other | 8 | 2.0 |

**Table 3. Uses of Mobile Payment Services**

### 5.2. Internal Consistency

All the Composite Reliability (CR) scores for the measurement model were found to be above the recommended value of 0.7 (Hair et al., 2014) which means that very high levels of internal consistency reliability were demonstrated among all reflective latent variables as shown in Figure 4.

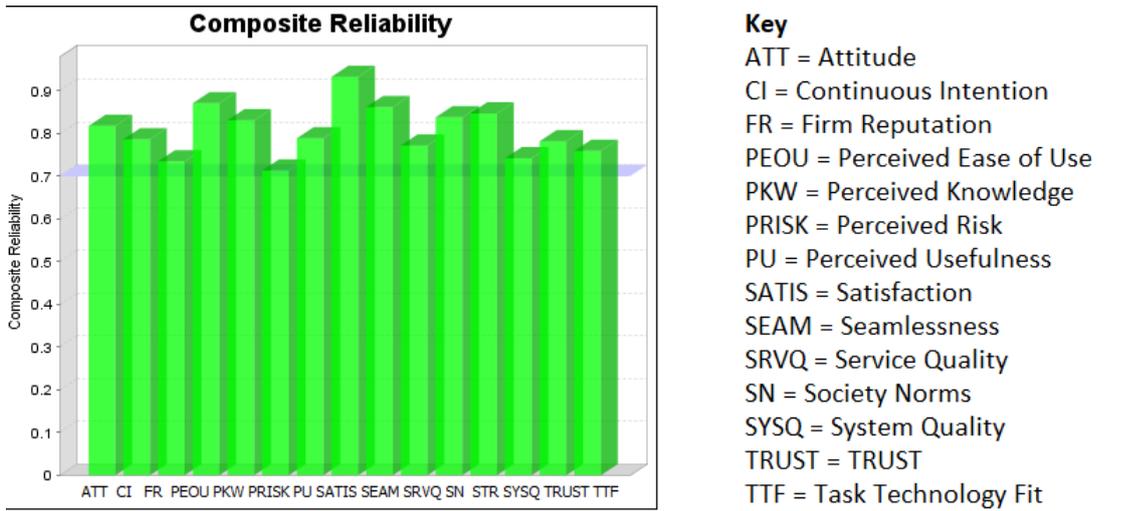

**Figure 4. Composite Reliability**

### 5.3. Assessing significance and Relevant of the Structural Model Relationships

Regression estimates (RE) and the p-value were used to explain relevance and significance, respectively. The study found that at 95% confidence interval, society norms (RE=0.153, p=0.003) and perceived usefulness (RE=0.159, p = 0.005) had the strongest effect on continuous intention at 15.3% and 15.9% of the total variance in continuous intention to use respectively and their paths were significant. Also, perceived ease of use (RE =0.126, p= 0.016) and attitude (RE=0.111, p = 0.048) had 12.6% and 11.1% of continuous intention to use respectively and their paths were significant. However, satisfaction (RE = 0.057, p = 0.243) and trust (RE = 0.020, p = 0.676) had low effect on continuous intention to use at 5.7% and 2% respectively and their paths were not significant at 95% confidence interval. Figure 5 shows the results of the structural model.

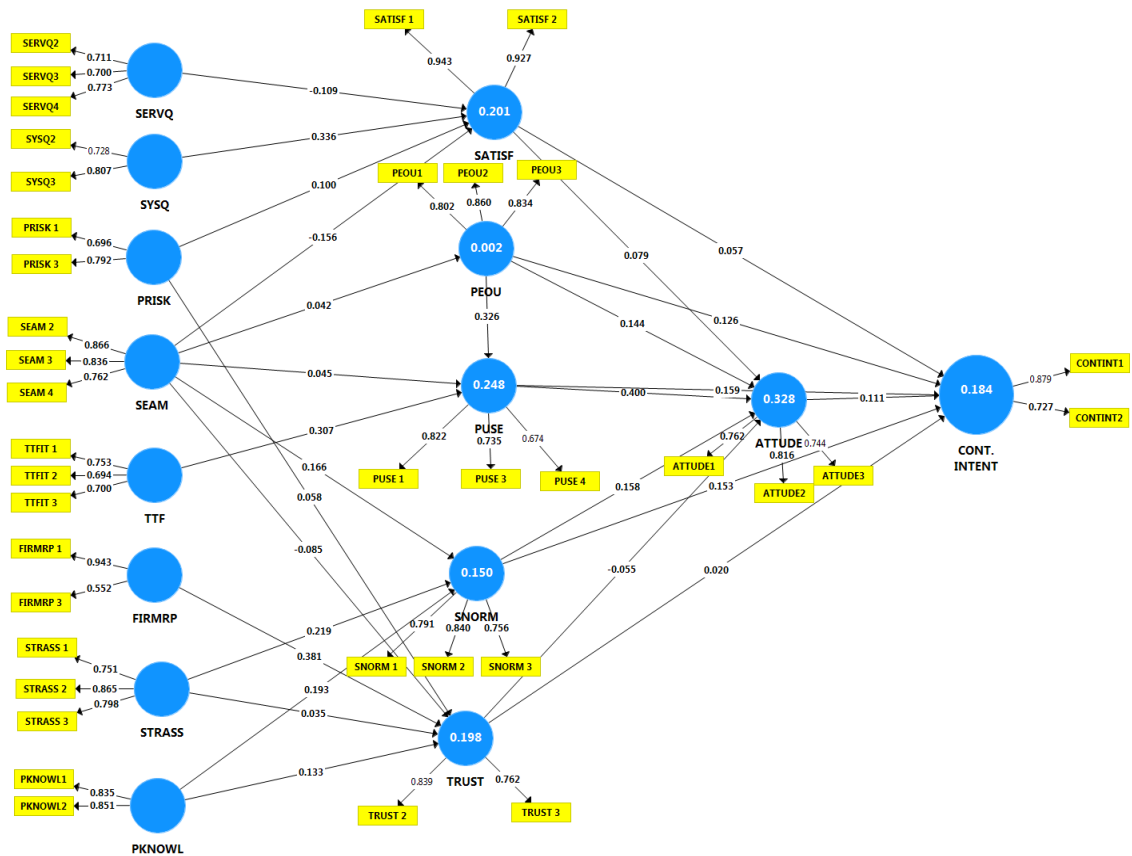

**Figure 5. Results of the Structural Model**

### 5.2.1. Satisfaction

Satisfaction was strongly explained by system quality at 33.6% (RE = 0.336, p = 0.000) followed by seamlessness at 15.6% (RE= - 0.156, p = 0.001). Both service quality (RE= - 0.109, p = 0.023) and seamlessness portrayed a negative relationship with satisfaction while their paths were both significant. This means that one unit drop in service quality and seamlessness leads to 10% and 15% loss of satisfaction respectively. Perceived risk (RE=0.10, p = 0.034) explained 10% of satisfaction and its path was significant.

The findings of this study have shown that there is no positive effect of satisfaction on continuous intention to use mobile payments (p = 0.770). This finding contrasts with Ahmed and Ali (2017) findings that satisfaction positively affected continuous intention to use mobile money transfer. However, most studies do not include satisfaction as of predictor of continuous intention or behaviour intention (Bradley, 2009; Lee et al., 2003).

### 5.2.2. Trust

The findings of this study has shown that trust was highly predicted by firm reputation (RE = 0.381, p = 0.000), then prior knowledge (RE = 0.133, p = 0.003) with both paths being significant. But worth of noting here is the negative relationship portrayed by seamlessness (RE = -0.85, p =0.064), although the path was found not to be significant. Perceived risk (RE=0.58, p = 0.08) and structural assurances (RE = 0.35, p = 0.476) had low trust and the paths were not significant.

This finding has shown that user's trust does not have a positive effect on continuous intention. This is similar to other mobile money transfer findings and mobile banking (Ahmed & Ali, 2017; Koenig-lewis, Palmer, & Moll, 2010). Perhaps, this finding may be explained well if we look at the predictors of trust. However, a possible explanation could be that trust is earned over a long period of time and therefore, mobile payments users might not have used the systems for a long time to know whether they have trust in the system.

### 5.2.3. Attitude

The finding of this study has shown that 40% of attitude variance was influenced by perceived usefulness (RE=0.400) and the path was significant (p=0.000). This is quite high considering the other constructs contributed less i.e. society norms at 15.8% (RE = 0.158), perceived ease of use at 14.4% (RE = 0.144), satisfaction at 7.9% (RE = 0.079), and trust at 5.5% (RE = - 0.055). The paths for society norms and perceived ease of use were both significant (p = 0.001 and p = 0.002 respectively) while the paths for satisfaction and trust were not significant (p = 0.770 and p = 0.203 respectively).

The attitude construct had one major hypothesis that attitude positively affects the continuous intention to use mobile payments. The estimates for H6 were $\beta = 0.111$, t-value = 1.99, and p-value = 0.048, which means that the hypothesis was supported. Therefore, the hypothesis was accepted.

### 5.2.4. Perceived Usefulness

Perceived usefulness was found to be strongly influenced by perceived ease of use (RE =0.326) followed by task technology fit (RE = 0.307), with both paths being statistically significant (p = 0.000). Seamlessness had only 4.5% (RE = 0.045) of perceived usefulness, and the path was found not to be significant (p = 0.312).

In this study, perceived usefulness was hypothesised to positively affect both attitude and continuous intention. It is implied from these results that the more useful people find a mobile payment system or service, the more likely they are to develop a positive attitude towards that system or service and, in turn, the more likely they will continue using that system or service.

### 5.2.5. Perceived Ease of Use

Perceived ease of use was hypothesised to positively influence continuous intention and perceived usefulness. Hypothesis H5c has already been discussed above. The estimates of H5a (β = 0.126, t-value = 2.42, p-value = 0.016) and H5b (β = 0.144, t-value =3.09, p-value = 0.002) supported the hypotheses and therefore both hypotheses were accepted.

### 5.2.6. Society Norms

Society norm was hypothesised to positively influence continuous intention and attitude. However, explaining society norms to understand what these results mean might be difficult if we do not explain and understand the predictors of society norms.

In this study, Society norms was found to be highly influenced by structural assurances (RE =0.219, p = 0.008) followed by prior knowledge (RE = 0.193, p = 0.000) then seamlessness (RE = 0.165, p =0.000) with all paths being significant.

## 6. DISCUSSION

The findings of this study have shown that young adults (57%) were the majority users of mobile payments. This could be because young people are the ones who are most likely to be technologically savvy and are happy to try new things. Furthermore, women (60.3%) used mobile payments more than men. It is most likely that most of these female users are still in school and, therefore, tend to use mobile payments to receive and send money from and to their family and friends. There was no big difference between those with lower education and those with higher education in terms of usage of mobile payments.

On mobile payment usage, the findings of this study were similar to the findings of the Reserve Bank of Malawi that many people use mobile payment services to send money to others (RBM, 2017). This could be because when mobile payment services were introduced in Malawi in 2011 by Airtel, the emphasis was on sending money before other services were incrementally introduced. The second most used mobile payment service was buying airtime (24%) for both telecom-led and bank-led services. This could be because telecommunication companies drive the mobile payment services agenda, and in their marketing, they just emphasise on buying airtime using one's mobile wallet. Surprisingly, just 9.4% use mobile payment services for buying goods in shops and stores. These findings are in contrast with the current trends in other African countries. For example, in Kenya it was reported that in 2016, the major uses of mobile payments (particularly mobile money) were paying and receiving salaries and bulk payment (90%), depositing money (85%), withdrawing money (98%), buying airtime (69%) and receiving remittances (64%) (The Economist, 2016). This shows that mobile payment users in Malawi are not using mobile payments to their fullest capacity. Therefore, the industry players and regulators should inform mobile payment users to leverage the services available to them.

### 6.1. Considerations for Mobile Payment Industry in Malawi

Based on the findings of this study, several considerations to the mobile payment industry have been proposed to help the industry reach out to its customer base and influence customers' continuous intention to use mobile payments in Malawi.

### 6.1.1. Improved business models

Perceived usefulness was found to be the most significant factor which influenced continuous intention to use mobile payments. Several studies in the mobile banking space had reported similar findings (Agwu, 2017; Ahmed & Ali, 2017). Significance in perceived usefulness was influenced by perceived ease of use and task technology fit. To boost people's perceived usefulness of mobile payments, the regulator could make a policy that every merchant must start accepting mobile payments. This includes paying for government services like road traffic services and fines, immigration services like passport application and renewal. This could make people see this befitting their tasks and lifestyle and increase their perception of mobile payment's usefulness. As it is now, most people in Malawi do not see how mobile payments fit into their daily tasks because it is not widely available and accepted. Changing this perception would improve people's perception of mobile payments.

### 6.1.2. Marketing strategies

Perceived ease of use of mobile payments was found to influence perceived usefulness, and ease of use was directly influenced by both attitude and continuous intention to use mobile payments. This finding is similar to other studies in Fintech and augmented reality (Chuang, Liu, & Kao, 2016; Chung, Han, & Joun, 2015). Therefore, marketing efforts by service providers should focus on showing people how easy it is to use mobile payments. This could be done by using visual arts like videos, comic stories, and even infographics. Current marketing efforts by service providers are customer acquisition-centric and not informative-centric. If industry players understand this, they could save money by speaking directly to the people's hearts and, in turn, changing their perceptions.

### 6.1.3. Customer engagement on security, privacy and transparency issues

Society norm was found to be highly significant in influencing continuous intention to use mobile payments. Other studies have found similar results (Ahmed & Ali, 2017). In this study, society norms were significantly influenced by structural assurances, prior knowledge and were negatively influenced by seamlessness. Structural assurances are more about customers feeling safe about their privacy and trust in the system in case of system failure and having no hidden costs. Banks and telecom companies should educate users about security threats of their funds and privacy of their transactions and themselves as individuals. For example, providing assurances that their funds are safe in terms of system failure or a possible hack. They could go further to have agreement statements during registration, thereby taking responsibility to refund people funds should such cases happen. This would create a generally positive image of mobile payments in the public domain, which will inform opinion about using mobile money since society norms influence continuous intention.

### 6.1.4. Interoperability of payment systems

Seamlessness was found to be highly significant in influencing society norms. This means that people value interoperability, and a lack of it leads to a negative image to the public. The need for more interoperability was also proposed and discussed by the global mobile telecommunication Companies' Association (GSMA) in their state of the industry 2014 report (GSMA, 2014). If mobile payments industry players want to reap the benefits of their investments, they need to start making their services seamless as this will help create confidence and trust among the general public and will lead to better attitude and continuous intention to use mobile payments.

## 7.    CONCLUSION

The findings of this study have suggested that for users to continuously use mobile payments society has a crucial role to play and the industry must put more efforts to understand their users. Enhancing structural assurances by improving privacy and guaranteeing safety of funds and making systems and services as seamless as possible will most likely positively improve society's attitude towards mobile payments and eventually leads to continuous intention to use a service. Therefore, based on society norms, what society thinks about a service has a significant impact on individual usage of that service. For mobile payments to succeed, service providers must do all they can to make sure that they are maintaining a positive image with society. In return, society norms may be influenced by the seamlessness of services and structural assurances.

In addition, when users find mobile payments to be useful, the more the users may continuously use mobile payments. Perceived usefulness also has great positive impact on user's attitudes. Therefore, service providers must create a need among the users so that they continue to find mobile payments useful and eventually positively affect their attitude and influence their continuous intention to use behaviour. Furthermore, the easier it is to use a mobile payments service, the more the users may find that service useful.

# REFERENCES AND CITATIONS